\documentstyle[aps,prl,twocolumn,epsfig]{revtex}

\catcode`\@=11
\def\abstract#1{\gdef\@abstract{
\begin{center}
\parbox{14cm}
{\small\rm\ignorespaces#1\par}
\end{center}}}
\def\@maketitle{%
\@preprint
\@title
\ifdim\prevdepth=-1000pt \prevdepth0pt\fi
\@authoraddress
\@date
\@abstract}
\def\section{\@startsection{section}{1}{\z@}{3.5ex plus 1ex minus .2ex}
{2.3ex plus .2ex}{\large\bf}}
\def\thesection{\arabic{section}.}

\def\appendix{\setcounter{section}{0}
\def\thesection{Appendix \Alph{section}:}
\def\theequation{\Alph{section}.\arabic{equation}}}
\def\@citex[#1]#2{\if@filesw\immediate\write\@auxout{\string\citation{#2}}\fi
  \def\@citea{}\@cite{\@for\@citeb:=#2\do
    {\@citea\def\@citea{,\penalty\@m}\@ifundefined
       {b@\@citeb}{{\bf ?}\@warning
       {Citation `\@citeb' on page \thepage \space undefined}}%
\hbox{\csname b@\@citeb\endcsname}}}{#1}}
\def\citer{\@ifnextchar [{\@tempswatrue\@citexr}{\@tempswafalse\@citexr[]}}
\def\@citexr[#1]#2{\if@filesw\immediate\write\@auxout{\string\citation{#2}}\fi
  \def\@citea{}\@cite{\@for\@citeb:=#2\do
    {\@citea\def\@citea{--\penalty\@m}\@ifundefined
       {b@\@citeb}{{\bf ?}\@warning
       {Citation `\@citeb' on page \thepage \space undefined}}%
\hbox{\csname b@\@citeb\endcsname}}}{#1}}
\relax

\begin{document}

\thispagestyle{empty}
\begin{center}

\parbox{17.5cm}
{
\begin{flushright}
CERN-TH/99-126\\
SLAC-PUB-8146\\
SHEP 99/04\\
hep-ph/9905312
\end{flushright}
\vspace*{1.5cm}
\begin{center}
{\Large\bf QCD Factorization for $B\to\pi\pi$ Decays:\\[0.2cm] 
Strong Phases and CP Violation in the Heavy Quark Limit${}^*$}\\
\vspace*{1.5cm}
\sc{M. Beneke}${}^a$, \sc{G. Buchalla}${}^a$,
\sc{M. Neubert}${}^b$, \sc{C.T. Sachrajda}${}^c$\\
\vspace*{1cm}
${}^a$Theory Division, CERN, CH-1211 Geneva 23, Switzerland

${}^b$Stanford Linear Accelerator Center,  
Stanford University, Stanford, CA 94309, USA

${}^c$Department of Physics and Astronomy, University of Southampton, 

Southampton SO17 1BJ, UK\\

\vspace*{1.5cm}
{\bf Abstract}\\
\end{center}
We show that, in the heavy quark limit, the hadronic matrix 
elements that enter $B$ meson decays into two light mesons can 
be computed from first principles, including `non-factorizable' 
strong interaction corrections, and expressed in terms of form 
factors and meson light-cone distribution amplitudes. The 
conventional factorization result follows in the limit when 
both power corrections in $1/m_b$ and radiative corrections 
in $\alpha_s$ are neglected. We compute the order-$\alpha_s$ 
corrections to the decays $B_d\to\pi^+\pi^-$, $B_d\to\pi^0\pi^0$ and
$B^+\to\pi^+\pi^0$ in the heavy quark limit and briefly discuss the 
phenomenological implications for the branching ratios, strong phases 
and CP violation.\\
\vskip-5pt PACS numbers: 12.38.Bx, 13.25.Hw\\[5.0cm]

{\small 
\noindent ${}^*$ The work of MB is supported in part by the 
EU Fourth Framework Programme `Training and Mobility of
Researchers', Network `Quantum Chromodynamics and the Deep Structure 
of Elementary Particles', contract FMRX-CT98-0194 (DG 12 - MIHT). 
The research of MN is supported by the Department of 
Energy under contract DE--AC03--76SF00515.
CTS acknowledges partial support from PPARC through grant GR/K55738.}}
\end{center}

\newpage
\thispagestyle{empty}
\parbox{17cm}
{\mbox{}}

\newpage
\thispagestyle{empty}
\parbox{17cm}
{\mbox{}}

\newpage
\setcounter{page}{0}

\title
{QCD Factorization for $B\to\pi\pi$ Decays:\\[0.0cm] 
Strong Phases and CP Violation in the Heavy Quark Limit}

\author{M. Beneke${}^a$, G. Buchalla${}^a$,
  M. Neubert${}^b$, C.T. Sachrajda${}^c$}
\address{${}^a$Theory Division, CERN, CH-1211 Geneva 23,
                Switzerland}
\address{${}^b$Stanford Linear Accelerator Center,  
Stanford University, Stanford, CA 94309, USA}
\address{${}^c$Department of Physics and Astronomy, University
  of Southampton, Southampton SO17 1BJ, UK}

\date{May, 11,  1999}

\abstract
{We show that, in the heavy quark limit, the hadronic matrix 
elements that enter $B$ meson decays into two light mesons can 
be computed from first principles, including `non-factorizable' 
strong interaction corrections, and expressed in terms of form 
factors and meson light-cone distribution amplitudes. The 
conventional factorization result follows in the limit when 
both power corrections in $1/m_b$ and radiative corrections 
in $\alpha_s$ are neglected. We compute the order-$\alpha_s$ 
corrections to the decays $B_d\to\pi^+\pi^-$, $B_d\to\pi^0\pi^0$ and
$B^+\to\pi^+\pi^0$ in the heavy quark limit and briefly discuss the 
phenomenological implications for the branching ratios, strong phases 
and CP violation.\\
\vskip-5pt PACS numbers: 12.38.Bx, 13.25.Hw\\[-1.2cm]}
\maketitle

\narrowtext

The detailed study of $B$ meson decays is a key source of information 
for understanding CP violation and the physics of flavour. The 
interest in this field is reinforced by the numerous
upcoming experiments that will test crucial
aspects of $B$ decay properties with unprecedented scope and
precision. Among the large number of $B$ decay channels, two-body
non-leptonic modes, such as $B\to\pi\pi$, $B\to\pi K$ etc.,
open a particularly rich field of phenomenological investigation.
A theoretical treatment, however, is generally
complicated owing to the non-trivial QCD dynamics related to the 
all-hadronic final state. 

In this Letter we describe important simplifications that occur in the 
limit $m_b\gg\Lambda_{QCD}$, when the $b$ quark mass is large 
compared to the strong interaction scale $\Lambda_{QCD}$. We find that 
in this limit the hadronic matrix elements for, say, $\bar B\to\pi\pi$ 
can be represented in the form 
\begin{eqnarray}
\label{fact1}
\langle \pi\pi|Q|\bar{B}\rangle  &=& \langle \pi|j_1|\bar{B}\rangle 
\langle \pi|j_2|0\rangle\cdot
\nonumber\\
&& \hspace*{-1cm}\cdot\Big[1+\sum r_n \alpha_s^n + {\cal O}
(\Lambda_{QCD}/m_b)\Big],
\end{eqnarray}
where $Q$ is a local operator in the weak effective Hamiltonian and 
$j_{1,2}$ are bilinear quark currents. Neglecting power corrections 
in $\Lambda_{QCD}$ {\em and} radiative corrections in $\alpha_s$, 
the original matrix element factorizes into a form factor times a 
decay constant (we call this conventional factorization). 
At higher order in $\alpha_s$ this simple factorization
is broken, but the corrections can be calculated systematically in 
terms of short-distance coefficients and meson light-cone 
distribution amplitudes. This is
similar in spirit to the well-known framework of perturbative
factorization for exclusive processes in QCD at large momentum
transfer \cite{LB}, as applied, for example, to 
the electromagnetic form factor of the pion.
An interesting consequence of (\ref{fact1}) is that strong interaction 
phases are formally of order $\alpha_s$ or 
$\Lambda_{QCD}/m_b$ in the heavy quark limit. If this limit 
works well, the approach discussed here allows us to calculate 
these phases systematically; CP violating weak 
phases can then be disentangled. Here we present a numerical analysis of 
$B\to\pi\pi$ decay amplitudes based on the heavy quark limit. We also 
briefly discuss important power corrections, which should eventually 
be estimated in order to obtain a satisfactory phenomenology at realistic 
$b$ quark masses.
Details of the argument that leads to the factorization formula
(\ref{fform}) below will be explained in a forthcoming paper.

The effective weak Hamiltonian describing 
$\bar B$ decays is given by \cite{BBL}
\begin{equation}
\label{heff}
{\cal H}_{eff}=
\frac{G_F}{\sqrt{2}}\sum_{p=u,c}\lambda_p
\bigg[ C_1 Q^p_1+C_2 Q^p_2+\!\!\sum_{i=3 \ldots 6,8}\!\!C_i Q_i\bigg],
\end{equation}
where $\lambda_p=V^*_{pd}V_{pb}$. The $Q_i$ are local
$\Delta B=1$, $\Delta S=0$ operators, and $C_i$ the corresponding
short-distance Wilson coefficients. We neglect electroweak penguin 
operators and all terms not relevant to $\bar{B}\to \pi\pi$ decays.

\begin{figure}[h]
   \vspace{-2.6cm}
   \hspace*{1cm}
   \epsfysize=24cm
   \epsfxsize=16cm
   \centerline{\epsffile{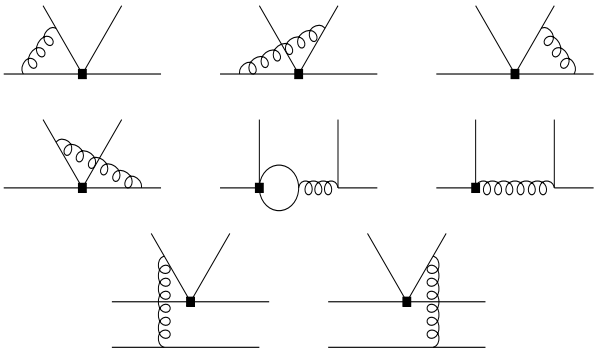}}
   \vspace*{-16.4cm}
\caption[dummy]{\small Order $\alpha_s$ corrections to the hard 
scattering kernels $T^I_i$ (first two rows) and $T^{II}_i$ 
(last row). In the case of $T^I_i$, the spectator quark does 
not participate in the hard interaction and is not drawn. 
The two lines directed upwards represent the two quarks that make up 
$\pi_2$.
\label{figlett1}}
\end{figure}

The essential theoretical problem for obtaining the $\bar B\to\pi\pi$
amplitudes is the evaluation of the hadronic matrix elements
$\langle\pi\pi|Q_i|\bar B\rangle$. Let $\pi_1$ denote the pion that picks 
up the light spectator quark in the $\bar{B}$ meson, and $\pi_2$ the 
pion whose valence partons are supplied by the weak decay of the 
$b$ quark. In the heavy quark limit both pions emerge with large 
energy $m_B/2$ (in the $\bar{B}$ rest frame). Power counting based 
on the asymptotic form of the leading-twist pion distribution amplitude 
shows that a leading-power contribution to the 
$\langle\pi\pi|Q_i|\bar B\rangle$ matrix element requires both valence 
quarks of $\pi_2$ to carry energy of order $m_b$. The $q\bar{q}$ pair 
is ejected from the weak interaction region as a small-size colour 
singlet object. As a consequence soft gluons with momentum of order 
$\Lambda_{QCD}$ decouple at leading order in $\Lambda_{QCD}/m_b$, 
and $\pi_2$ can be represented by its leading-twist light-cone distribution 
amplitude. On the other hand, the spectator quark in the 
$\bar{B}$ meson carries momentum of order $\Lambda_{QCD}$ and 
is transferred as a soft quark to $\pi_1$, unless it undergoes a 
hard interaction. The endpoint suppression of the pion wave function 
is not sufficient to ensure the dominance of hard interactions. 
[We adopt the point of view that for realistic $b$ quark masses 
perturbative Sudakov suppression does not cut off soft contributions 
efficiently enough before one enters the non-perturbative regime.] 
Therefore $\pi_1$ cannot always be represented by its light-cone distribution 
amplitude. At leading power in $\Lambda_{QCD}/m_b$, we find that the 
soft interactions can be absorbed into the $B\to \pi_1$ form factor. 
On the other hand, any interaction of the spectator quark 
with the quarks of $\pi_2$ is hard at leading power
and can be written as a convolution 
of three light-cone distribution amplitudes. This discussion can be 
summarized by the factorization formula
\begin{eqnarray}\label{fform}
&&\langle\pi(p')\pi(q)|Q_i|\bar B(p)\rangle =
f^{B\to\pi}(q^2)\int^1_0 dx\, T^{I}_i(x)\Phi_\pi(x)
 \nonumber \\
&& \ \ +\int^1_0 d\xi dx dy\, T^{II}_i(\xi,x,y)\Phi_B(\xi)
\Phi_\pi(x) \Phi_\pi(y),
\end{eqnarray}
which is valid up to corrections of relative order 
$\Lambda_{QCD}/m_b$.
Here $f^{B\to\pi}(q^2)$ is a $B\to\pi$ form factor evaluated
at $q^2=m^2_\pi\approx 0$, and $\Phi_\pi$ ($\Phi_B$) are 
leading-twist light-cone
distribution amplitudes of the pion ($B$ meson),
normalized to 1. The $T^{I,II}_i$ denote hard-scattering kernels, 
which are calculable in perturbation theory. $T^{I}_i$ starts 
at ${\cal O}(\alpha^0_s)$; at higher order 
in $\alpha_s$ it contains `non-factorizable'
gluon exchange, including penguin topologies, see the first two rows 
of Fig.~\ref{figlett1} for the corrections at order $\alpha_s$. 
Hard, `non-factorizable' interactions involving the spectator 
quark are part of $T^{II}_i$ (last row of Fig.~\ref{figlett1}). 
Annihilation topologies also exist, but are power-suppressed 
in $\Lambda_{QCD}/m_b$.
The significance of the factorization formula is that all the
non-perturbative effects in the $B\to\pi\pi$ amplitudes can be absorbed
into the form factor and the light-cone wave functions.

The following comments are in order:

(i) When $\alpha_s$ corrections are neglected $T_i^{II}$ is 
zero and $T_i^{I}$ is an $x$-independent constant. 
Conventional 
factorization in terms of the form factor and the pion decay constant
is then recovered as a rigorous prediction in the infinite 
quark mass limit. The perturbative corrections are process-dependent, 
but calculable. Their inclusion cancels the scale-dependence 
of the leading-order factorization result. 

(ii) The infrared finiteness of the hard scattering amplitude follows 
because the infrared divergences in the first four diagrams of 
Fig.~\ref{figlett1} cancel in their sum. This cancellation is the 
technical manifestation of Bjorken's colour transparency argument 
\cite{BJ}. Colour transparency does not apply to hard gluon 
interactions. These, however, are suppressed by $\alpha_s$ and are
calculable.

(iii) The hard scattering contribution to the $B\to\pi$ form factor 
is suppressed by one power of $\alpha_s$ relative to the soft 
contribution, in which the $B$ meson spectator undergoes no hard 
interaction. As a consequence the assumption that $B\to\pi\pi$ 
can be treated entirely in the hard scattering picture of \cite{LB} 
would miss the leading contribution in the heavy quark limit.

(iv) The decay amplitude acquires an imaginary part through the 
hard scattering kernels. In the heavy quark limit, the strong interaction 
phases can therefore be computed as expansions in $\alpha_s$. In 
terms of hadronic intermediate states that saturate the unitarity 
relation this implies systematic cancellations among many intermediate 
states with potentially large individual rescattering phases. An 
estimate of rescattering effects on the basis of Regge theory is 
not compatible with the picture that emerges in the heavy quark limit.

(v) The factorization formula (\ref{fform}) generalizes to the decays 
into a heavy-light final state, if the heavy particle absorbs the 
$B$ meson spectator quark. In this case the second line in (\ref{fform}) 
is power-suppressed and only the form factor term survives. An expression 
of this form has been used by Politzer and Wise to compute the 
1-loop corrections to the decay rate ratio $\Gamma(\bar{B}\to D^*\pi)/
\Gamma(\bar{B}\to D\pi)$ \cite{PW}. 
The factorization formula does {\em not} hold for heavy-light final states, 
in which the light meson absorbs the $B$ meson spectator quark, or for 
a heavy-heavy final state. In this case, conventional factorization 
can also not be justified.

The result of an explicit calculation of the $\bar B\to\pi\pi$
decay amplitudes at order $\alpha_s$ can
be compactly expressed as
$\langle\pi\pi|{\cal H}_{eff}|\bar B\rangle=
G_F/\sqrt{2}\sum_{p=u,c}\lambda_p\langle\pi\pi|{\cal T}_p|\bar B\rangle$,
where
\begin{eqnarray}\label{tp}
{\cal T}_p &=& 
a^p_1(\pi\pi) \,(\bar ub)_{V-A}\otimes (\bar du)_{V-A}
\nonumber \\
&+& a^p_2(\pi\pi) \,(\bar db)_{V-A}\otimes (\bar uu)_{V-A} 
\nonumber \\
&+& a_3(\pi\pi) \,(\bar db)_{V-A}\otimes (\bar qq)_{V-A}\nonumber \\
&+& a^p_4(\pi\pi) \,(\bar qb)_{V-A}\otimes (\bar dq)_{V-A} \nonumber \\
&+& a_5(\pi\pi) \,(\bar db)_{V-A}\otimes (\bar qq)_{V+A}\nonumber \\
&+& a^p_6(\pi\pi) \,(-2)(\bar qb)_{S-P}\otimes (\bar dq)_{S+P}.
\end{eqnarray} 
The symbol $\otimes$ is defined through 
$\langle \pi\pi|j_1\otimes j_2|\bar{B}\rangle\equiv 
\langle \pi|j_1|\bar{B}\rangle \langle \pi|j_2|0\rangle$. 
A summation over $q=u,d$ is implied. 
Note that the term proportional to $a_6^p(\pi\pi)$ 
results in a power correction that should be dropped in the 
heavy quark limit. We will comment further on this term below.

Together with $a^c_1(\pi\pi) = a^c_2(\pi\pi) =0$ and the leading-order 
coefficient $a_6^p(\pi\pi)=C_6+C_5/N$, the 
QCD coefficients $a^p_i(\pi\pi)$ read at next-to-leading order (NLO) 
\begin{eqnarray}
a^u_1(\pi\pi) &=& C_1+\frac{1}{N}C_2 +
\frac{\alpha_s}{4\pi}\frac{C_F}{N}\,C_2\,F, 
\label{a1u}\\
a^u_2(\pi\pi) &=& C_2+\frac{1}{N}C_1 +
\frac{\alpha_s}{4\pi}\frac{C_F}{N}\,C_1\,F, 
\label{a2u}\\
a_3(\pi\pi) &=& C_3+\frac{1}{N}C_4 +
\frac{\alpha_s}{4\pi}\frac{C_F}{N}\,C_4\,F, 
\label{a3}\\
a^p_4(\pi\pi) &=& C_4+\frac{1}{N}C_3 -\frac{\alpha_s}{4\pi}\frac{C_F}{N}
\biggl[\bigg(
\frac{4}{3}C_1+\frac{44}{3}C_3
\nonumber\\
&&\hspace*{-1.2cm}+\frac{4f}{3}(C_4+C_6)\bigg)
\ln\frac{\mu}{m_b}
+\left(G_\pi(s_p)-\frac{2}{3}\right)C_1
\nonumber\\
&&\hspace*{-1.2cm}+\!\left(G_\pi(0)+G_\pi(1)-f^I_\pi-f^{II}_{\pi}+
\frac{50}{3}\right)C_3
\\
&&\hspace*{-1.2cm}+\Big(3 G_\pi(0)+G_\pi(s_c)+G_\pi(1)\Big)(C_4+C_6)
+G_{\pi,8} C_8\biggr],
\nonumber\\
a_5(\pi\pi) &=& C_5+\frac{1}{N}C_6 +
\frac{\alpha_s}{4\pi}\frac{C_F}{N}\,C_6\,(-F-12). 
\label{a5}
\end{eqnarray}
Here $C_F=(N^2-1)/(2N)$ and $N=3$ ($f=5$) is the number
of colours (flavours). [Note that our definition of $C_1$ and $C_2$ differs 
from the convention of \cite{BBL}, where the labels $1$ and $2$ are 
interchanged.] The internal quark mass in penguin diagrams
enters as $s_p$, where $s_u=0$ and $s_c=m^2_c/m^2_b$.
In addition we have used ($\bar x\equiv 1-x$)
\begin{eqnarray}\label{g8pi}
&&F=-12\ln\frac{\mu}{m_b}-18+f^I_\pi+f^{II}_\pi,
\\
&&f^I_\pi=\int^1_0 \!\!\!dx\,g(x)\Phi_\pi(x),\quad
G_{\pi,8}=\int^1_0 \!\!\!dx\, G_8(x)\Phi_\pi(x),
\\
&&G_\pi(s)=\int^1_0 \!\!\!dx\,G(s,x)\Phi_\pi(x),
\end{eqnarray}
with the hard-scattering functions
\begin{eqnarray}
\label{g8x}
&&g(x)=3\frac{1-2x}{1-x}\ln x-3 i\pi,
\quad G_8(x)=\frac{2}{\bar x},
\\
\label{ggsx}
&&G(s,x)=-4\int^1_0  \!\!\!du\,u(1-u)\ln(s-u(1-u)\bar x-i\epsilon).
\end{eqnarray}
The hard spectator scattering contribution is given by
\begin{equation}\label{hbpi}
f^{II}_{\pi}=\frac{4\pi^2}{N}\frac{f_\pi f_B}{f_+(0) m^2_B}
\int^1_0\!\!\!d\xi\,\frac{\Phi_B(\xi)}{\xi}
\left[\,\int^1_0\!\!\!dx\,\frac{\Phi_\pi(x)}{x}\right]^2,
\end{equation}
where $f_\pi$ ($f_B$) is the pion ($B$ meson) decay constant,
$m_B$ the $B$ meson mass, $f_+(0)$ the $B\to\pi$ form factor
at zero momentum transfer, and $\xi$ the light-cone momentum fraction
of the spectator in the $B$ meson.
$f^{II}_{\pi}$ depends on the wave function
$\Phi_B$ through the integral $\int^1_0d\xi \,\Phi_B(\xi)/\xi
\equiv m_B/\lambda_B$. This introduces one new hadronic parameter
$\lambda_B$. Since $\Phi_B(\xi)$ has support only for $\xi$
of order $\Lambda_{QCD}/m_B$, $\lambda_B$ is of order $\Lambda_{QCD}$.

Writing the transition operator ${\cal T}_p$ in terms
of the QCD coefficients $a^p_i(\pi\pi)$ is a convenient notation
for phenomenological applications. The notation generalizes the 
conventional parameters $a_{1,2}$ \cite{NS}, 
which are seen to be process-dependent 
beyond leading order. We emphasize that in the
present context the $a^p_i(\pi\pi)$ are not phenomenological parameters,
but genuine predictions of QCD in the heavy quark limit. 
The Wilson coefficients $C_i$ entering
the $a^p_i(\pi\pi)$ are to be taken at NLO \cite{BBL}, where we
consistently drop terms of ${\cal O}(\alpha^2_s)$ 
in (\ref{a1u})--(\ref{a5}). The physical amplitudes
derived from (\ref{tp}) are independent of the renormalization 
scale ($\mu$) and scheme through ${\cal O}(\alpha_s)$. 
The coefficients $a_1(\pi\pi)$--$a_5(\pi\pi)$ multiply scale and
scheme independent matrix elements of (axial-)vector currents. 
Accordingly for $a_1(\pi\pi)$--$a_5(\pi\pi)$ the scale and 
scheme dependence in the
Wilson coefficients $C_i$ is canceled by the ${\cal O}(\alpha_s)$
corrections in the hard-scattering amplitudes.
In the case of $a^p_6(\pi\pi)$, a scale and scheme dependence remains,
which is precisely the one needed to cancel the corresponding
dependence in the matrix elements of the (pseudo-)scalar currents,
multiplying $a^p_6(\pi\pi)$ in (\ref{tp}).   
Besides the $\ln(\mu/m_b)$ terms the hard-scattering
amplitudes contain a scheme dependent constant, which we have obtained
in the NDR scheme as defined in \cite{BJLW}. This fixes the scheme
to be used for the NLO coefficients $C_i$.

At NLO the factorization coefficients $a^p_i(\pi\pi)$ acquire
complex phases, entering through the functions $g(x)$ and $G(s,x)$
in (\ref{g8x}) and (\ref{ggsx}). Being of order $\alpha_s$, these
phases are generically small, except in cases where the lowest order
contribution is numerically suppressed. This happens {\it e.g.\/}
for $a^u_2(\pi\pi)$. Physically, the phases arise 
from final state rescattering, which is due to hard
gluon exchange, and hence perturbative, in the heavy quark limit.
The generation of strong interaction phases through the penguin function
$G(s,x)$ has been discussed many years ago \cite{BSS} and is
commonly referred to as the Bander--Silverman--Soni (BSS) mechanism.
In the present approach, the gluon virtuality 
$k^2=\bar x m^2_B$ in the
penguin diagram, which has usually been treated as a
free phenomenological parameter, has a well-defined
meaning. The $x$-dependence
of $G(s,x)$ is convoluted with the pion wave function
$\Phi_\pi(x)$, leaving no ambiguity as to the value of $k^2$.
In addition we identify a further source of rescattering phases, 
represented by the function $g(x)$. This effect
corresponds to hard gluon exchange between the two outgoing pions.
Together with the BSS mechanism, it accounts for the
complete asymptotic rescattering phases in $\bar B\to\pi\pi$ in
the heavy quark limit.

Another novel result is the
existence of the contribution from hard scattering 
involving the spectator quark in the $B$ meson, expressed by
$f^{II}_{\pi}$ in (\ref{hbpi}). This mechanism is completely missed
in phenomenological models of factorization. It 
is particularly important for the small
coefficient $a^u_2(\pi\pi)$, where it leads to a sizable
enhancement. Using $f_\pi=131\,$MeV, $f_B=(180\pm 20)\,$MeV, 
$f_+(0)=0.275\pm0.025$, $\lambda_B=0.3\,$GeV and the 
asymptotic pion wave function $\Phi_{\pi}(x)=6 x\bar{x}$, we find 
$f^{II}_{\pi}\approx 6.4$. The  poor knowledge of the $B$ meson 
parameter $\lambda_B$ makes this number rather uncertain.

Numerical values for the $a^p_i(\pi\pi)$ are shown in Table \ref{tab:ai}, 
using the pole masses $m_b=4.8\,$GeV, $m_c=1.4\,$ 
GeV, the $\overline{\rm MS}$ masses $\bar m_t(\bar m_t)=167\,$GeV,  
$(\bar m_u+\bar m_d)(2\,\mbox{GeV})=9\,$MeV and 
$\Lambda^{(5)}_{\overline{MS}}=225\,$MeV as input parameters. 
$a_6^p(\pi\pi)$ multiplies the  
$\Lambda_{QCD}/m_b$-suppressed, but chirally enhanced combination
\begin{equation}\label{chi12}
r_\chi=\frac{2m^2_{\pi^+}}{\bar m_b(\mu)(\bar m_u(\mu)+\bar m_d(\mu))}
\approx 1.18 \quad[\mbox{at }\mu=m_b].
\end{equation}
In the following analysis, we give two results, one neglecting 
$a_6^p(\pi\pi)$ as formally power-suppressed, the other keeping 
the leading-order expression for $a_6^p(\pi\pi)$.

It is now straightforward to evaluate
the decay amplitudes and branching ratios. The latter are given by
$Br(\bar B\to\pi\pi)=\tau_B/(16\pi m_B)\cdot|A(\bar B\to\pi\pi)|^2 S$, 
where $S=1$ for $\pi\pi=\pi^+\pi^-$, $\pi^-\pi^0$ and $S=1/2$ for
$\pi\pi=\pi^0\pi^0$. $\tau_B$ denotes the $B$ meson lifetimes: 
$\tau(B^+)=1.65\,$ps, $\tau(B_d)=1.56\,$ps.
The decay amplitudes read
\begin{eqnarray}\label{abpm}
&&A(\bar B_d\to\pi^+\pi^-) = i\frac{G_F}{\sqrt{2}}m^2_B f_+(0)f_\pi
|\lambda_c|\cdot \\
&&\hspace*{0.5cm}\cdot\bigl[R_b e^{-i\gamma}
\left(a^u_1(\pi\pi)+a^u_4(\pi\pi)+a^u_6(\pi\pi)r_\chi\right)
\nonumber\\
&&\hspace*{0.8cm}-\left(a^c_4(\pi\pi)+a^c_6(\pi\pi)r_\chi\right)\bigr],
\nonumber\\
\label{ab00}
&&A(\bar B_d\to\pi^0\pi^0) = i\frac{G_F}{\sqrt{2}}m^2_B f_+(0)f_\pi
|\lambda_c|\cdot \\
&&\hspace*{0.5cm}\cdot\bigl[R_b e^{-i\gamma}
\left(-a^u_2(\pi\pi)+a^u_4(\pi\pi)+a^u_6(\pi\pi)r_\chi\right)
\nonumber\\
&&\hspace*{0.8cm}-\left(a^c_4(\pi\pi)+a^c_6(\pi\pi)r_\chi\right)\bigr],
\nonumber\\
\label{abm0}
&&A(B^-\to\pi^-\pi^0) = i\frac{G_F}{\sqrt{2}}m^2_B f_+(0)f_\pi
|\lambda_c|\cdot\\
&&\hspace*{0.5cm}\cdot \,R_b/\sqrt{2}\cdot e^{-i\gamma}
\left(a^u_1(\pi\pi)+a^u_2(\pi\pi)\right).\nonumber 
\end{eqnarray}
Here $R_b=(1-\lambda^2/2)|V_{ub}/V_{cb}|/\lambda$, where $\lambda=0.22$
is the sine of the Cabibbo angle, $\gamma$ is the phase of $V^*_{ub}$,
and we will use $|V_{cb}|=0.039\pm 0.002$, $|V_{ub}/V_{cb}|= 
0.085\pm 0.020$. We find 
the branching fractions 
\begin{eqnarray}\label{brpi}
&&Br(\bar B_d\to\pi^+\pi^-) = 
\nonumber\\
&&\hspace*{0.5cm} 6.5\,[6.1]\cdot 10^{-6}
\left|\,e^{-i\gamma} + 0.09\,[0.18] \,e^{i\cdot 12.7\,[6.7]^\circ}\,
\right|^{\,2}\!\!,
\\
&&Br(\bar B_d\to\pi^0\pi^0) = \nonumber\\
&&\hspace*{0.5cm} 5.2\,[7.7]\cdot 10^{-8}
\left|\,e^{-i\gamma} + 0.73\,[1.11] \,e^{-i\cdot 137\,[149]^\circ}\,
\right|^{\,2}\!\!,
\\
&&Br(B^-\to\pi^-\pi^0) = 4.3\,[4.3]\cdot 10^{-6},
\end{eqnarray}
where the default values correspond to neglecting $a_6^p(\pi\pi)$ and 
the values in brackets use $a_6^p(\pi\pi)$ at leading order. While the 
predictions for the $\pi^+\pi^-$ and $\pi^-\pi^0$ final states are 
relatively robust, with errors on the order of $\pm 30\%$ due to the 
input parameters, the decay into $\pi^0\pi^0$ depends very sensitively 
on the parameter $\lambda_B$ that controls the hard spectator 
scattering. If it is significantly smaller than $0.3\,$GeV, 
a branching fraction of order $10^{-6}$ cannot be excluded. 
Eq.~(\ref{brpi}) can be converted into a result for the time-dependent 
CP asymmetry as a function of the CKM angle $\alpha$.
The direct CP asymmetry in the $\pi^+\pi^-$ mode is
approximately $4\%\cdot\sin\gamma$.

The approach discussed here allows us to formulate, for the first
time, rigorous predictions of QCD for exclusive non-leptonic
$B$ decays in the heavy quark limit.
On the other hand, as the dependence on the formally power-suppressed 
coefficient $a_6^p(\pi\pi)$ demonstrates, the asymptotic limit may 
be problematic at $m_b\approx 5\,$GeV and the applicability of the 
theory has to be decided on a case-by-case basis. The most important 
power corrections are those that depend on the chirally enhanced 
combination (\ref{chi12}). The $\alpha_s$ corrections to all such terms 
can in fact be identified. However, the factorization formula breaks 
down in this case, because the relevant twist-3 wave functions 
do not fall off fast enough at the endpoints. A detailed discussion 
of this point, and of its phenomenological consequences, 
will be given elsewhere.

\vspace*{-0.3cm}


\begin{table}
\caption{The QCD coefficients $a^p_i(\pi\pi)$ at NLO for three different 
renormalization scales $\mu$. Leading order values are shown
in parenthesis for comparison.}
\label{tab:ai}
\begin{center}
\begin{tabular}{lccc}
 & $\mu=m_b/2$ & $\mu=m_b$ & $\mu=2 m_b$ \\ \hline
$a^u_1(\pi\pi)$ & 
     $1.047+0.033i$ & $1.038+0.018i$ & $1.027+0.010i$ \\
 & $(1.038)$ & $(1.020)$ & $(1.010)$ \\
$a^u_2(\pi\pi)$ & 
     $0.061-0.106i$ & $0.082-0.080i$ & $0.108-0.064i$ \\
 & $(0.066)$ & $(0.140)$ & $(0.200)$ \\
$a_3(\pi\pi)$ & 
     $0.005+0.003i$ & $0.004+0.002i$ & $0.003+0.001i$ \\
 & $(0.004)$ & $(0.002)$ & $(0.001)$ \\
$a^u_4(\pi\pi)$ & 
     $-0.030-0.019i$ & $-0.029-0.015i$ & $-0.026-0.013i$ \\
 & $(-0.027)$ & $(-0.020)$ & $(-0.014)$ \\
$a^c_4(\pi\pi)$ & 
     $-0.038-0.009i$ & $-0.034-0.008i$ & $-0.031-0.007i$ \\
 & $(-0.027)$ & $(-0.020)$ & $(-0.014)$ \\
$a_5(\pi\pi)$ & 
     $-0.006-0.004i$ & $-0.005-0.002i$ & $-0.003-0.001i$ \\
 & $(-0.005)$ & $(-0.002)$ & $(-0.001)$ \\
$a^{p}_6(\pi\pi)r_\chi$ & --
     & -- & --\\
 & $(-0.036)$ & $(-0.030)$ & $(-0.024)$ \\
\end{tabular}
\end{center}
\end{table}


\begin{thebibliography}{99}
\bibitem{LB}
\vspace*{-1cm}
G.P. Lepage and S.J. Brodsky, Phys. Rev. {\bf D22}, 2157 (1980).
\bibitem{BBL}
G. Buchalla, A.J. Buras and M.E. Lautenbacher,
Rev. Mod. Phys. {\bf 68}, 1125 (1996). 
\bibitem{BJ}
J.D. Bjorken, Nucl. Phys. (Proc. Suppl.) {\bf B11}, 325 (1989).
\bibitem{PW}
H.D. Politzer and M.B. Wise, Phys. Lett. {\bf B257}, 399 (1991).
\bibitem{NS}
M. Neubert and B. Stech, hep-ph/9705292.
\bibitem{BJLW}
A.J. Buras et al., Nucl. Phys. {\bf B400}, 37 (1993). 
\bibitem{BSS}
M. Bander, D. Silverman and A. Soni, 
Phys. Rev. Lett. {\bf 43}, 242 (1979).
\end{thebibliography}
\end{document}